# Raising Awareness of Conveyed Personality In Social Media Traces

Bin Xu[2], Liang Gou[1], Anbang Xu[1],
Jalal Mahmud[1], Dan Cosley[2]
[1]IBM Almaden Research Center
[2]Cornell University

**ABSTRACT**

Users' persistent social media contents like posts on Face-book Timeline are presented as an "exhibition" about the person to others, and managing these exhibitional contents for impression management needs intentional and manual ef-forts. To raise awareness of and facilitate impression man-agement around past contents, we developed a prototype called PersonalityInsight. The system employs computa-tional psycho-linguistic analysis to help users visualize the way their past text posts might convey impressions of their personality and allowed users to modify their posts based on these visualizations. We conducted a user study to evaluate the design; users overall found that such a tool raised aware-ness of the fact and the ways personality might be conveyed through their past content as one aspect of impression man-agement, but that it needs design improvement to offer action-able suggestions for content modification, as well as careful thinking about impression management as one of many val-ues people have about their digital past.

**ACM Classification Keywords**
H.5.m. Information Interfaces and Presentation (e.g. HCI): Miscellaneous

**INTRODUCTION**
A key concern people have when they post contents to digi-tal media, whether text posts, pictures, videos, or comments, is impression management: creating an image of themselves they want others to see. Goffman's frontstage/backstage metaphor, which casts our behavior as a public perfor-mance [12], is often used to think about the choices people make at the time they post in social media [35].

Unlike a stage performance, however, the data people post to social media generally persists. As Hogan points out, these data can be repurposed by systems, presented as an "exhibi-tion" about a person to other people [16]. For example, the Facebook Timeline presents a record of your past interactions with the system, and this record, too, can create impressions of yourselves. Sometimes the original poster has little con-trol over the system and people try to balance their concerns about the image presented with goals of personal recordkeep-ing [34].

However, managing these exhibitional contents for impres-sion management is not easy. Again taking Facebook Time-line for example, some users, especially active ones who have a lot of posts on timeline, might find that manually reviewing each post frequently, changing their access settings or delet-ing them takes tremendous time and efforts. Further, they might not be aware of the impressions they are creating, or of the audiences who might be seeing them.

Our research goal is to design tools to help users manage these contents based on the presentational effects of their ex-hibitional contents in social media. In this paper, we take a step toward this goal through presenting and evaluating a system designed to help people be aware of the images their past contents create, understand better how particular pieces of the contents contribute to their images, and easily modify the contents or access to them.

The system takes text posts in the Facebook Timeline, as the platform of our target exhibitional online contents, and we take the personality as the lens of impression management and self-presentation. Specifically, we use the Big-five per-sonality traits in the system, which are often a part of how people think about their own and others' images (e.g., "he's introvert"). Using an computational personality model as the back-end, we developed a prototype called, PersonalityIn-sights, to explore what happens when a system present peo-ple with information about their personality derived from that their writings. The system gives three different levels of Big-five traits presentation: the overall level, which presents what Big-five traits the whole timeline present; the individual post level which shows how each individual post contributes to the overall Big-five image; and the word level which shows how each word in a post associates with each trait in Big-five. The system also provides content management features to let users manage their Timeline contents.

We conducted a user study on this system to analyze the ef-fects of explicit presenting and explaining the derived on-line personality images, and evaluated the designs of such a facilitating tool. Our study shows that after using the sys-tem, participants raised their awareness of the presentational effects of Facebook Timeline contents. Some of the users raises their concerns and self-consciousness for future self-disclosure, though not all of them. Users also gave feedbacks about how different levels of analytic results given by the sys-tem have different values and limitations to facilitate their content management. They overall found a tool like this is useful for the awareness and reflection of their current per-sonality presented on Facebook Timeline, but that it lacks the ability to offer actionable suggestions that people might take—and that, as also found by Zhao et al., people balancing impression management concerns with a desire to retain per-sonal pasts [34, 35]. We then discussed how these findings can be used for the design of future tools like this aiming to help people understand and manage their online images that the digital contents conveys to others.

## BACKGROUND AND RELATED WORK

### Impression Management and Online Exhibitional Content

People are socially aware of how they are perceived and eval-uate by others in a social settings, and attempt to convey images that are associated with their attainment of desired goals [19]. This process of individuals' efforts to control the impression formed by others is referred as impression man-agement. People manage their impressions to others in dif-ferent ways. For example, some people do cosmetics, di-ets and even plastic surgery to be more physically attractive; politicians practice speeches to be more persuasive. Besides these physical appearances and public image management, our daily, mundane interactions with others also serve as self-presentation and leave impressions to others [11, 12]. Goff-man developed a dramaturgical metaphor in his theory to il-lustrate the embedded self-presentation in our daily interac-tion, and users the term "performance" to refer to the activi-ties of a person to others in a social setting [12].

The same impression management practices also extend to social media platforms. Studies have recognized that one of the most important factors affecting people's online self-disclosure is impression management [2, 3, 7, 8, 24]. And despite being developed for offline interaction contexts, Goff-man's dramaturgical approach can be also applied in social network sites and social media studies to understand users' impression management practices[2, 3, 7].

However, different from offline interaction, online interac-tions on social system often leave digital traces even after the interaction is finished. Hogan emphasized this distinction and extended Goffman's theory by using "exhibition" to present the digital contents that are generated from "performance" but stay persistent in the system and visible in the future[16].

Managing these exhibitional contents is not easy work. Xuan and Lindley pointed out that the curation of these contents need users' intentional actions, but that current tools don't support this kind of curation well [34]. Part of the problem is that tools don't support the curation process well: privacy settings are often hard to use, and active users need to sort through large volumes of content [34].

| | Definitions |
|---|---|
| Openness | The extent to which a person is open to experience a variety of activities. |
| Conscientiousness | A tendency that a person acts in an organized or spontaneous way. |
| Agreeableness | A tendency to be compassionate and cooperative towards others. |
| Extraversion | A tendency to seek stimulation in the company of others. |
| Neuroticism | The extent to which a person's emo-tion is sensitive to the environment |

Table 1: Definitions of Big Five Model of personality traits

We also see these tools as failing to support curation goals around impression management. Facebook's "others see me as" feature allows people to see themselves as others see them, at least through the Facebook Timeline, and in prin-ciple, this could be used to support impression management of past content. As a tool for helping people think about their image, however, this view is not very helpful. The system provides neither interpretation nor guidance, which are often important in tools designed to help people reflect on their be-havior [20]. In particular, it doesn't help people think about what might be important to consider in their image, nor spe-cific ways their content might contribute to that image.

### Personality as a Meaningful, Computable Lens

Impression is a complex term, and a person's activity could present several impression-related traits about herself, such as values, emotions, interests, beliefs, and so on [19].

In this paper, we use big-five personality traits as the main in-dicator of image. We choose personality because it has been recognized as one of the most focused and concerned facts in online self-disclosure on social network sites [7, 21], and several factors of users' personality have been found to ei-ther associate or predict some social media usage including impression management concerns [1, 5, 18, 26]. Following the choice of recent work around personality [5, 18, 26] we adopt the Big-Five, a descriptive model of personality with five main factors: Conscientiousness, Agreeableness, Neu-roticism, Extraversion, and Openness [23]. Each Big Five factor represents a collection of personality traits. Table 1 shows the definitions of these factors [23].

To support personality presentation management, we first need to obtain the personality traits from users' exhibi-tional contents. One possible way is to use a crowdsourc-ing method [17] such as send the user's online contents to crowdsourcing platform and let the crowd rate how the user's personality is. This method has some advantages, notably around processing non-text content and incorporating actual human reactions. However, it also raises privacy issues: shar-ing the contents to strangers could be a privacy violation to the user; and hiding or eliminating identity-related con-tents means losing some personality presentational contents as well. A second way is to let the user's network rate for her personality. However, since the interaction between a user and her contacts could also occur beyond one platform or on-line, the rated personality traits could also be influenced by other channels other than the online contents [33]. Also, both approaches are hard to scale up.

Fortunately, with the development of psycho-linguistic and text mining analysis methods, there are multiple computational methods to model people's personality from their digital footprints. The seminal works from Golbeck et al. use psycholinguistic analysis to model Big 5 personality from Twitter [13]. Holtgraves found correlations between Linguistic Inquiry and Word Count (LIWC) categories in users' text messages and their personality traits [27]. Yarkoni also found significant correlation among Big 5 personality and LIWC with blog data [32]. Mairesse et.al illustrated the usefulness of different algorithms for automatic recognition of personal-ity traits both in conversation and text [22]. Gou el.al studied feasibility and effectiveness of modeling users' personality traits with online social media contents [15].

Not only is personality computable, but there is initial evidence that it is useful and interesting to users. For example, Wang et.al developed a tool to visualize different personal-ity dimensions and showed that computed personality anal-ysis effectively supports various user privacy configuration tasks [28]. Warshaw developed a mobile app to understand people's reactions to computed personality profile [30].

Previous studies like [15, 28, 30] focused on the represen-tation of the personality traits, and studied how users would disclose these visualized traits to others. However, in this study, our goal is about online self-presentation management. First, we want to investigate how people react to the results of personality traits of their own, not others, and secondly, we aim at providing clues to help users get insights on how their traits look like in certain way, rather than just presenting the computed levels of the traits [15, 28].

**Research Aims**
The aim of this study is to investigate whether computational modeling of personality techniques can be used as a facilita-tion method for self-presentation management on online so-cial media. We attempted to achieve this goal through a sys-tem probe in which people use a prototype to experience the technology, observing their reactions on the results that come from the analytic algorithms and asking them about their im-pressions. We chose this methodology because without the actual implementation of the algorithm, it is difficult to in-vestigate the reactions and feedbacks of the users to such a design. In this next section we discuss our design choice to meet this goal in our prototype system, and present the hy-potheses and study design to test the hypotheses.

**SYSTEM AND HYPOTHESES**
Figure 1 shows the overview architecture of PersonalityIn-sights. First, the system uses Facebook Graph API to get the current user's own posts from her Timeline, because our re-search scope focuses on the projected online image by a per-son's own exhibitional content. Then, the posts are passed to a Big5 modeling module to generate a personality profile including the five OCEAN factors (Openness, Conscientious-ness, Extraversion, Agreeableness, and Neuroticism). These traits are computed at the overall level (including all posts from a user) and the individual post level. Then, the sys-tem offers visual representations of the overall image, and the contributions of each individual post towards the overall im-age. Users can then see which words and posts most strongly contribute to their profile. Finally, users can manage their contents by editing the content, changing the visibility to dif-ferent audience, or deleting posts.

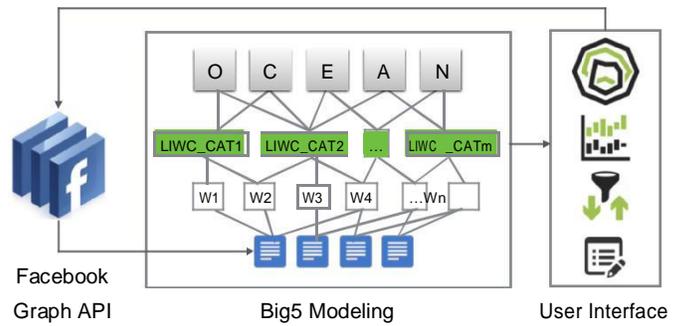

Figure 1: The system architecture of PersonalityInsights.

**Big Five Modeling**
A key component of the system is to compute a user's personality profile from her posts. The Big5 personality model is adopted from previous work [15, 32]. It is build upon the psycho-linguistic analysis by computing the correlation between personality traits and the usage of certain words or word categories in people's writing. The model uses the LIWC (Linguistic Inquiry and Word Count) dictionary [25] in which word categories are defined to capture peoples so-cial and psychological states. In this model, posts are first decomposed as a bag of words with standard text processing techniques. A personality trait is computed by a linear com-bination of relevant LIWC category scores and each LIWC category score is the normalized frequency of the words that belong to the LIWC category and are also used in the posts. The final trait score is then normalized and converted into percentile score over a large trait score pool we built with 3 millions social media users.

For users' understanding and management of their contents, this module also provides relevant insights of how a trait is de-rived from their posts. For example, in the model, the "Agree-ableness" factor is positively correlated to LIWC category of "Inclusive" which includes words such as "us", "with", and "along". The model includes both LIWC category scores and matched words along with the trait scores.

**User Interface**
The computed overall profile and relevant insights are then presented to users with our interactive visual interface, with which users can explore, comprehend and manage their on-line personality presence derived from the contents.

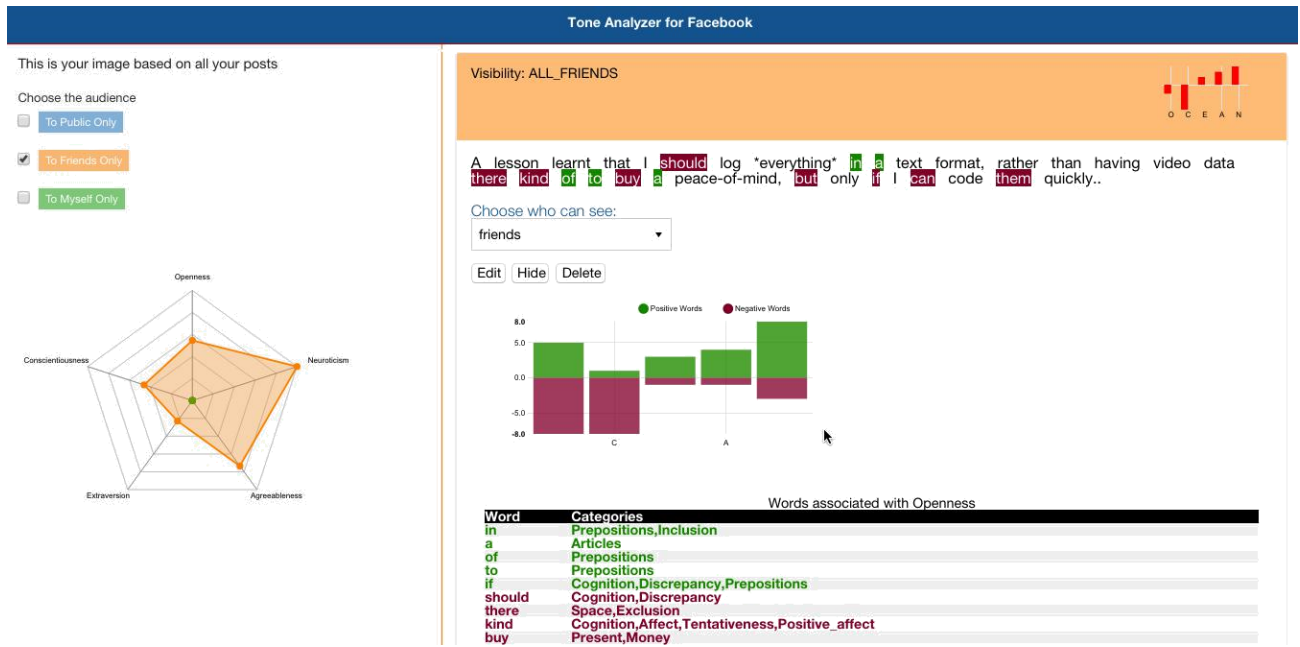

Figure 2: The PersonalityInsights interface. The radar chat shows the overall personality based on the whole posts set. On the right, it shows each post's analytic results. The upper right chat on shows how this post contributes to each trait towards the overall personality profile. A user can also change the audience of each post, and edit, hide or delete a post. The middle bar chart shows numbers of positively and negatively associated words in the post to each trait. When the user click each bar, a table is shown below the chart, presenting all words associated with traits. Green highlighted words are positively associated, while the red highlighted are negatively associated.

One goal of the system is to offer users an overall view of their personality profile, depicted in the radar chart in Fig-ure 2. This view is derived from the user's own Facebook Timeline contents fetched at signin, excluding posts uploaded by others (but shown in the user's timeline, for instance, if they are tagged in the post). Three overall personality profiles are computed and visualized based on the audience of public, friends and myself (private posts). Checking/unchecking the option boxes allows users to compare the profile they present to different audiences.

Another goal is to help people make sense of the profile by understanding how the individual post is contributing to the overall personality image. To achieve this, the interface pro-vides a post-level analysis that shows the contribution of each post towards the traits in the overall image through a summary bar chart in the upper right corner. It also uses a larger bar chart under each post and separates the total contribution of the post into positive and negative contributions toward each of the five traits (the bar charts in Figure 2. For this insight, the system processes the words in each post, and calculates the accumulated weights of each word's association to each Big5 trait by using the Big5 modeling module.

The system also gives a word-level insight that shows all the words in a post associated to each trait in big five. The sys-tem presents the weights of positive or negative correlation, and also the explanations on why the words are positively or negatively associated with the traits from our Big5 model.

The explanations are based on the words' LIWC categories that are associated to personality traits in our algorithm for the overall personality computation, since LIWC categories give users a social and psychological summarization of words used in the post.

Last, the system should also empower users to manage the contents after users perceive the results and the cues to un-derstand the results. As shown in Figure 2, we provide some affordances to interact with the post as same as Facebook Timeline does. Users can edit a post, hide it or delete it. When they edit a post, both the post-level and word-level personality analysis are updated in real-time, and the overall image will be updated only after users finish the editing. To reduce the workload for users with a large number of posts, the interface also provides a sorting function to help users to identify posts that make relatively large contributions to either the positive or negative computation of traits.

**Hypotheses**

In summary, the current design of system includes two ana-lytic features to users: first, it provides a personality profile computed by the system. Secondly, it provides visual cues to help users make sense how the results are derived.

We hypothesize several influences of these features. First, showing both the personality profile and the cues on how the posts and words can reveal their personality might let users start to realize that the contents, especially some linguistic cues, could reveal some of their personality. Previous stud-ies have use linguistic cues to recognize users' personality traits, for example, [28], however, normal users neither ac-knowledge the possibility of extracting personality from their textual contents, nor have a sense of how the extraction can be done. We hypothesize that:

H1.a. Seeing the personality profile will raise their awareness that their social media contents have a presentational function to show their personality;

H1.b. Seeing the cues of how the posts and words contributes to their personality profile will raise their awareness their lin-guistic use of certain words is at least one factor in their per-sonality formation online.

Secondly, if the computed personality profile does not match the user's own perceived personality profile, it might trigger concerns about their content that, in turn, might trigger con-tent management behaviors to bring them more into line.

H2. It might raise the user's concerns about her projected personality were not the desired one she wants to;

H3. The concerns eventually leads to their actions to change the contents based on the profile and the cues.

**Study Design**
We designed a lab study to examine the hypotheses, evaluate our design choices in the system, and also to get insights for future design and improvement.

Overview and participants
The overview of the study is as follows. The participants were invited to the lab, and first signed the consent form, and answer a pre-study survey. Then they were assigned to use both the PersonalityInsights interface and a parallel con-trolled interface described below, in counterbalanced order. Instructions were given before each session, telling them to review their Timeline posts as long as they want, and make as many or few changes as they want to the contents. We did not explicitly give any pre-defined task as Wang et.al did was because we aimed at investigating their management prac-tices in a general and nature setting, rather than a designated context. After finishing two sessions, they answered the post-study survey, and were interviewed by one of our researchers.

We recruited our participants from the Facebook Group of a large U.S. technology enterprise in California. A total of 16 users who had enough Timeline content to compute mean-ingful personality profiles (which based on accuracy results from prior work we defined as containing 1000 words) par-ticipated; they were paid by lunch vouchers after finishing the study.

Method and metrics
To test the hypotheses H1 and H2, we asked questions about users' awareness of contents' personality presentation effects, and their concerns about personality presentation, in both the pre and post-study surveys.

Because most users have probably not thought about the way past content might present a personality image [14], we thought it likely that people would report increases in aware-ness and concerns if they just used the PersonalityInsights interface—but that this effect might be simply due to the fact of thinking about personality, rather than effects of the actual data presented. To help tease apart the priming of personal-ity from the effect of the interface, we developed a control interface similar to that shown earlier in Figure 2. It did not present any analysis, but did present explanations about Big5 and access to the content management features, as shown in Figure 3.

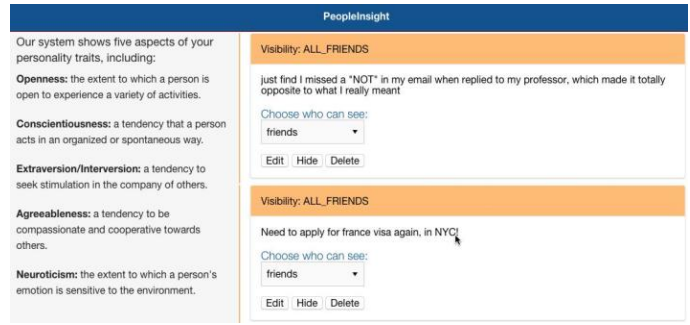

Figure 3: The Non-Insight interface. In this interface, the sys-tem will not show the analytic results nor the cues about the results, but it still has the instructions about what Big5 per-sonality model is, and also the content management features including editing, deletion and hiding the posts.

Each participant used both Non-Insight interface and the Per-sonalityInsights interface. For each interface, the system retrieves half of the participant's Facebook Timeline posts based on the posts' created time, and either generates the per-sonality analytic results in the PersonalityInsights interface, or simply shows the posts in the Non-Insight interface. The study sessions are counterbalanced by the order of using two interfaces, and the order of which half of the posts being used, and therefore we have 4 (2 interfaces order X 2 Facebook Timeline posts selection).

Because in this within-subjects design both interfaces might have had effects on awareness and content, the post-survey questions about awareness and content asked about the per-ceived effect of each interface separately.

We also chose not to collect "ground truth" personality mea-sures with scales like NEO Personality Inventory Revised [6] because we thought the concerns described in H2 would be caused by differences between the user's desired image and the computed one, rather than differences between their ac-tual image and the computed one.

To test hypothesis H3, we asked in the pre-study survey about their general use frequency of Facebook posting, and their general practices of reviewing and managing previous posts on Facebook Timeline; and then in the post-study survey we asked about their anticipated future practices on review-ing and content managing on Facebook Timeline in order to track changes in their actions. Because of the within-subjects design, in the post-study survey we asked how much they thought using each interface contributed to their future an-ticipated practices.

To round out the data collection, we asked demographic survey questions pre-study, and questions about overall system usability, usefulness, and likelihood of using such a system in the future post-study. We also collected log data on how often people performed content management in both interfaces. Finally, at the end of the study we conducted a short semi-structured interview asking participants how they perceived the personality results, whether and how each kind of insight result influenced their views and management of their Timeline posts, and any concerns and issues they had with the interfaces.

**RESULTS**

We now discuss each of the research question in turn: to what extent did the system raise awareness of the personality image conveyed by past content, raise concerns about that image, and lead to content management behaviors around it.

**"Now I Am Aware"**

To answer H1.a and H1.b, we compared how participants rated their awareness of personality and image in past content in the pre-study vs. post-study surveys. As shown in Fig-ure 4(a), after participating the study, participants were more aware of personality presentational effects of their Facebook Timeline contents. Further, this was not just because of in-voking the concept of personality; the Non-Insight interface had low ratings regarding change of awareness, while the Per-sonalityInsights interface had much higher ratings, as shown in Figure 4 (b).

Interview data also showed raised awareness of the idea that past content might convey aspects of their personality. Many participants did not know that there is a way to estimate their personality based on their Facebook timeline contents, and seeing the computed results is interesting to them:

"I haven't seen anything like this interesting analytic about my Facebook though (P05)"

The different levels of granularity (overall, posts, and words) provided different ways of reflecting on their presented personality. Overall image did lead some participants to think about what personality traits they wanted to convey:

"I did not know that on Facebook I am like not that agreeable to others, and now I know it, and I will post less posts that are not agreeable and try to be more corporative to others, in the future.(P04)"

For the post level results, participants found the results also helped them to get aware of how each post could contribute to the overall personality image:

"[The post level results] are useful, and make me aware of the personality can be learned from the posts (P02)"

Further, interface click log data showed that all participants used the sorting buttons to sort the posts, focusing on the posts that made the strongest contributions to the computed person-ality, and doing this in order to reduce the effort required:

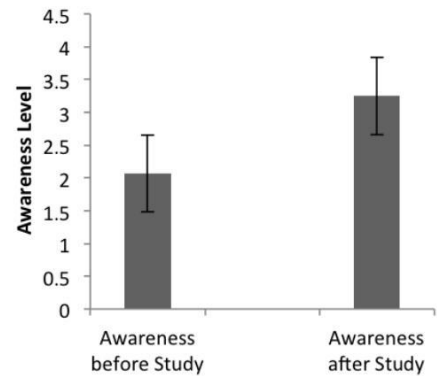

(a)

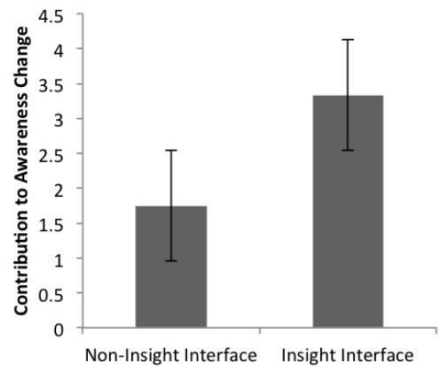

(b)

Figure 4: (a) The raised awareness of personality presentational effects of online exhibitional contents. Awareness level is from 1 to 5, where 1 means "not at all aware" and 5 means "extremely aware". $F(1, 15)=27, p<0.005$. (b) How the Non-Insight interface and PersonalityInsights interface influenced awareness changes. Influence level is from 1 to 5, where 1 is "not at all influential" and 5 means "extremely influential". $F(1, 15)=19.56, p<0.001$.

"Because I do not want to review all the posts of the timeline, but the PersonalityInsights interface ranks posts according to the personalities, and I can review the top ones, and the bottom ones. (P08)"

Participants used word level insights to try to understand why a given post was contributing to the overal computed person-ality, although most participants found it difficult to under-stand the word-level insights at beginning:

"The word level is a little bit confusing, because I think most of the words, the extracted positive and negative words are, like prepositions, articles that are less with particular meanings (P04)".

Although the word-level insights are not easy to understand, they still lead to a raised participants' awareness of how the use of words would show their personality. For example, P11 said "It is interesting to see that these words can show my personality, even though most time they do not have have the semantic meaning, like the 'a', 'one', how would they tell personality? I do not understand the categories, but I think it should be based on some sort of psychological reasons?" After looking at several posts, participants they also started to learn to make sense of the word-level results: "[it is] interesting, it makes me sounds very sensitive on this one because I say the word 'pretty' (P05)" .

**Conflicts Influences Concerns**

Our second hypothesis is that the conflicts between system analyzed personality traits and users' self-perceived personality would lead to concern about their self presentation by the exhibitional contents. As with awareness, concerns about the presentation of one's personality also emerge from the data. In the survey, as shown in Figure 5(a), the concern raises af-ter participants finished the experiment, and as shown in Fig-ure 5(b), the effect is not just about the awareness of person-ality, but driven more directly by the information presented in the PersonalityInsights interface.

Interestingly, these concerns may have been less about whether the computed personality presented a desired image and more about whether the system "got it right" in the sense of being an accurate model. On a survey question that asked about how well the computed personality agreed with their own perceptions of their personality, the average rating was a neutral 3 on a 5 point scale. In the interviews, participants also focused on how the computed personality agreed with those perceptions, rather than with a desired image:

"I think somewhat it is true. Like I am very open to new ideas. And some posts [analysis] have very high accuracy, like the conscientiousness, the posts ranked at the bottom of the con-scientiousness are the posts that I complained that I don't have time for my exams. I think that has very high accuracy. (P06)"

**Awareness, Concern, but no Curation**

This absence of a desired image may have affected people's desire to curate their past content. Log data shows that participants did not make many changes to these posts as we hy-pothesized. Only few participants changed the posts (average edit action per user =1.1, SE=1.45).

Still, despite the low frequency of intentional actions and the forces leading toward a lack of content management in the study, some participants reported that seeing the computa-tional personality analysis might influence future disclosure behavior for new content:

"First, I will be more careful posting things on Facebook. I do not want to post the too negative posts. The other thing is that now I know it could be posted to public and friends, and I want to make it more clear whether to public or to friends (P15)".

When we made our design choices, we aimed at designing and developing a tool that would eventually facilitate the exhibitional content management for impression management . We use Big5 personality presentation as the observation and Facebook Timeline as the exhibitional content sources, and though users did raise their awareness and concerns about ex-hibitional contents' personality presentation function, it is un-clear that the current design directly led to users' intentional actions of editing, deleting or hiding the posts. What does this result mean for the current design? Is a presentation fa-cilitating tool still necessary? If yes, what do our results tell us about the next design? In the next section we will discuss these questions, then we present some implications based on the discuss and some future works.

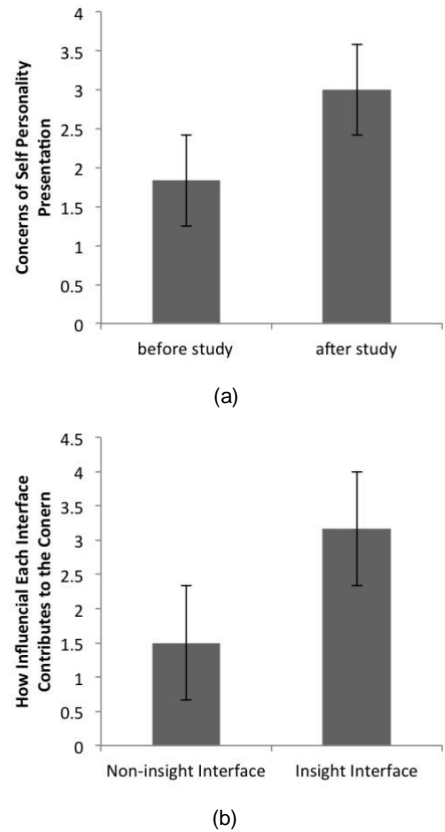

Figure 5: a) The raised concerns of personality presentational effects of online exhibitional contents. Concern level is from 1 to 5, where 1 means "not at all concerned " and 5 means "extremely concerned". F=9.14, p<0.01. (b) How the Non-Insight interface and PersonalityInsights interface influenced participant's change of concerns of presentational effects of their Facebook Timeline contents. Influence level is from 1 to 5, where 1 is "not at all influential" and 5 means "extremely influential". F(1, 15)=19.64, p<0.01.

**DISCUSSION**

**Interpreting Beyond Presenting**

We made one design choice of directly presenting the LIWC categories that the algorithm uses to compute the personality profile because we did not know what kind of explanations would be effective, and we want to probe the possibility with the most direct choice we have, the cues from the algorithm itself. In this sense, the explanations we give in the post-level cues and the word-level cues are more toward presenting the facts and leave the interpreting work to the users.

Although people got some value from the cues, they were not necessarily actionable; first because the presenting facts based on the LIWC categories did not provide direct suggestions and instructions on how to change the post: "But when I see the word insights, I don't know how to revise it. (P11)" This suggests that a system's role in such a facilitation is not only an information provider, but should also serve as an active agent that processes the facts to be more interpretable. One possible answer might be, as this participant suggested, to show alternate wordings that would convey different images: "I think if it could give suggestions of how the wording should be used, after this version, could be better. (P2)" This solution might make the user's job more like making choices rather than interpreting, and thus might lead to lower cognitive load.

But does this mean that the system should just provide the suggestions, or even more automatically, modify the contents on behalf of the users? From our results, it is implied that even without the actionable suggestions, letting users know how the system or algorithm generates the results, or simply put, what is happening in the system or algorithm did help users understand the computed results a little better. A study on machine translation mediated collaboration found more concreted evidences that showing the flaw, the alternatives of a machine translation system leads to better grounding [10], and better user experiences [31]. When, and how the infor-mation about the algorithms, here the LIWC cues, the posts' contribution, while in other systems for example Facebook, could be how NewsFeed algorithms filters posts, how "Peo-ple you may know" picks up candidates should be presented to users, and what impacts it has to users' reflection of their own behavior, their mental model of the system, and their re-actions to the awareness, are questions this study alone can not answer. But referring to the role of transparency in phys-ical systems like enterprise, governments, and communities, reveling certain transparency on the internal mechanism of a digital social system is a design question that needs explo-ration.

System Accuracy and Attribution
Questions about the computed model's fidelity also led some participants to discount the insights given by the system: "It is just like, oh I just see, it's another tool, that says something about yourself, and probably not very true. I think the best judge is your self (P08)". The attribution of system sometime also means sense-making of how the system did wrong. For example, "that might be not accurate, ..., I don't post many post ". Users also talked about how their certain actions would also the system to get the inaccurate results. For example, "I think because I changed my privacy settings once, and the system only get my old public posts, so I think that is why my public image looks different (P12)".

Another extended question is that beside the inaccuracy the model has, what other psychological factors that influences users' attribution to they system, or themselves. For example, theories on conformity emphasize that when people are presented by arguments, facts or conclusions that contract to their own, they experience an intension between their own psychological states (like perceptions, beliefs and et.al) and the external ones [4]. This intension also happened in the use of our system. The quote of P08's was an example. How to design the system to deal with these intensions, and why to design in a certain way are also needed insights for similar systems that might present some conflicting information with users'.

**Multiple Purposes and Practices of Exhibitional Contents**
Participants also reported that they would not hide or delete posts because the content itself and its meaning to them was more important in an exhibition context than the personality it might convey.

"I think when I posted some contents, I was not aware that on Facebook I post mostly to public or to friends, maybe because I do not use very frequently. But for posts to my friends, I'd like to post something happened in my life, to my friends. But for some interesting posts I see on other website, like some videos on Youtube, I'd like to share to public. And for myself, the posts could be like, I see something, and I want to record, but I don't have a system to record it, so i post on Facebook to myself (P04)".

This description aligns closely with ideas from Zhao et al. (2013) about the use of social media as a personal archive—even if that past self is different from the present.

"I don't want to revise my timeline very largely, because yeah, I'd like to see how my past, like my old me behave in a few years ago. Because at least I can see that at that period time, what my personality is, and maybe it is interesting to see how the personality is changing over time. (P05)"

Some participants also worried that sudden changes to old ex-hibitional contents might lead others to think the personality presented on Facebook is less real or consistent to the user's real personality offline, and thus has a risk being perceived as deceptive.

"I don't think I will use something like that, because it will make it fake to you, and if that is the way, I will just post to myself. I don't want to share such a thing with others (P14)".

Another reason is that the impression management is not the only practice users would act upon these contents. For ex-ample, when being asked about their general contents man-agement on Facebook Timeline, they mentioned about the audience would impact their posting behavior on Facebook Timeline.

"I think that depends on who is following me, because cur-rently I am still a student, and my only boss is my advisor, and she doesn't follow her student on Facebook, but when I graduate, and when I am at work, and my colleagues are fol-lowing me, and my bosses are following me, then I will be very careful for posting things on Facebook....then I will use this system more frequently (P16)".

Participants also mentioned about their content management focus on privacy: "No, I just think this can have some privacy issues, not personality issues. By privacy issues, I mean that I want to post something here but not there, that is the infor-mation that I am concern, but not personality...Maybe in the future I will consider this. I am not sure, I am think maybe I will post something like fake words to show my personality good, you know. (P06)"

This suggests that the "exhibition" has multiple purpose, which consists with the motivation in the use of Facebook [7], and impression management is only one of them. In turn, this raises the question about how to make the presentation facil-itation work along with other purpose and practices. Xuan et.al pointed out that on Facebook, users manage their Time-line while keep the purposes of performing, exhibiting and archiving at the same time while perceiving the platform af-fordance and leveraging the features together [35]. Wang et.al developed a privacy nudge while users post contents on Facebook, which implies an integrated privacy management aligned with self-disclosure behaviors [29].

In our system, the integration with other purposes is not salient by design, which was intentionally chosen by us in order to evaluate the personality management alone. How-ever, the results suggest that when we design a system, even though we aim at facilitating one purpose, it is important to understand how the purposes are combined with or even con-flicted to each other, to guide the design choices.

**Short-Term Awareness, Long Term Change?**
One value of raising the awareness that words, linguistic cues can also represent personality is direct users to interpret their image based on the contents when the future audiences are hard to predict, and the lost contexts are missed for review-ers out of the original contexts. But would this awareness lead to some useful actions that would let users manage their contents better? Unfortunately this study did not show. How-ever, we argue that the current design of the system eventually make the personality management more "salient" to users. Similar to the previous studies that also making privacy man-agement and Newsfeed filtering more salient [9] , where they found making these practices salient have long-term impacts on people's sharing behavior, which might also imply that for personality/impression management we also need a longitu-dinal study to find out what the raise awareness influences in a long run.

**CONCLUSION**
In this study, we designed and developed a proto-type,PersonalityInsights, to present our idea of using com-putational method to raise the awareness of, offer useful in-formation towards, and facilitate users' impression manage-ment around contents they have posted to social media in the past. The user study showed that although the system did in-crease awareness, more work will be needed to help users un-derstand how impressions are conveyed and what actionable strategies need to be taken for improving their images. More generally, it showed that impression management is one of many concerns people have when they are working with their past contents: that fidelity, consistency, respect for the past, and personal archiving are all values that coexist with impres-sion management. Future systems in this space will need to account for these issues.